\documentclass[11pt,a4paper]{article}

\usepackage[pdftex]{graphicx}
\usepackage{amsmath}
\usepackage{amssymb}
\usepackage{setspace} 
\usepackage{subfigure}
\begin{document}
\title{Precise Particle Tracking Against a Complicated Background: Polynomial Fitting with Gaussian Weight}
\author{Salman S. Rogers\thanks{salman.rogers@physics.org}, Thomas A. Waigh, Xiubo Zhao, Jian R. Lu \\
School of Physics and Astronomy, University of Manchester, Manchester M60~1QD, UK}
\maketitle

\section*{Abstract}
We present a new particle tracking software algorithm designed to accurately track the motion of low-contrast particles against a background with large variations in light levels. The method is based on a polynomial fit of the intensity around each feature point, weighted by a Gaussian function of the distance from the centre, and is especially suitable for tracking endogeneous particles in the cell, imaged with bright field, phase contrast or fluorescence optical microscopy. Furthermore, the method can simultaneously track particles of all different sizes, and allows significant freedom in their shape. The algorithm is evaluated using the quantitative measures of accuracy and precision of previous authors, using simulated images at variable signal-to-noise ratios. To these we add a new test of the error due to a non-uniform background. Finally the tracking of particles in real cell images is demonstrated. The method is made freely available for non-commencial use as a software package with a graphical user-inferface, which can be run within the Matlab programming environment.


\section{Introduction}
Particle tracking---the use of computer analysis to measure the motion of individual particles in a sequence of images---is critical for a number of biophysical techniques, including particle tracking microrheology \cite{Tseng2002,Weihs2006,waigh2006}, magnetic \cite{Bausch1998} and optical tweezers \cite{Ashkin1997}. It is also useful for studying the transport of fluorescently-labelled single-molecules and organelles within the cell \cite{Courty2006a,Watanabe2007a}, and is widely used for particle imaging velocimetry \cite{adrian2005,grant1997}.

A wide range of methods have been used to track particles in optical microscope images \cite{Miura2005,Cheezum2001,crocker1996grier,Sbalzarini2005,Carter2005}. The most accurate for tracking small spherical particles (i.e.~of diameter less than the wavelength of light) are based on the fitting of each intensity peak in the image with a two-dimensional Gaussian function. This method works well in images with a uniform background, but tends to fail if the background is complicated, or when the particles are larger or non-spherical \cite{Cheezum2001}. For larger or non-spherical particles, tracking methods based on calculating the brightness-weighted centroid of a particle tend to give good results \cite{Cheezum2001,crocker1996grier}. Here pixels are normally judged to be ``inside'' a particle if their brightness values are above a chosen threshold. However the latter method has an inherent drawback: that if a pixel on the edge of a particle is just above threshold in one frame, and just below threshold in the next, then a spurious jump in the centroid's position is recorded. This error becomes larger with increasing noise or decreasing diameter of the tracked particle. Methods based on calculating centroids also tend to fail if the image background is not uniform, since it becomes more difficult to exclude the background simply by subtraction or thresholding.

To date, the clearest analysis of errors in these tracking methods has been given by Cheezum et al.~\cite{Cheezum2001}, who consider numerically simulated images of a single particle in a microscope image. The simulation is simple but realistic, and the authors find that the signal to noise ratio $S/N$ is crucial in determining the tracking error. The error can be discussed in terms of \emph{accuracy} and \emph{precision}, which here respectively describe the \emph{determinate} error---i.e.~the bias $B$ in the position with respect to the geometry of the system, and the \emph{indeterminate} error---i.e.~the standard deviation $\sigma$ between measured and true positions. It is unfortunate that most scientists still rely on crude measures of their particle tracking accuracy or precision, such as measuring the apparent displacements of immobilised particles. This measure can only yield a lower bound to the error, as it ignores errors due to the displacement of the particle relative to the optical elements, CCD pixels, and visible features within the sample. Moreover, it is important to understand that tracking error is not analogous to the Rayleigh criterion, i.e.~it is not a constant measured in nanometers for the particular microscope configuration. Instead it depends strongly on the signal intensity, that is, the contrast between the particle and its surroundings. 

We present a new tracking method built on a different concept: a two-dimensional polynomial of 4th order is fitted to each particle image. The fit is weighted by a Gaussian function that is centred on the particle, with a decay length set to the particle's radius, which is measured from the image. The particle centre is itself determined by the polynomial's local maximum. This ``polynomial-fit Gaussian weight'' (PFGW) method is explained in detail below, but we can first see three obvious advantages of using a polynomial fit. Firstly, like Gaussian fitting, the fit is not sensitive to the edges of a particle. Secondly, like the centroid methods, the fit has significant flexibility towards the shape of a particle. Thirdly, unlike the previous methods, the procedure of determining the particle centre is inherently insensitive to the image background.

We evaluate the new fitting method using the analysis of Cheezum et al.~\cite{Cheezum2001}, and describe a new test of the effect of a simple non-uniform background on particle tracking errors. Finally we demonstrate the new method on real cell images where its accuracy and precision can also be appreciated.

The PFGW method is made freely available for non-commencial use, in the form of a software package with a graphical user-inferface, named PolyParticleTracker. The package requires the Matlab programming environment, version 7.0 or higher.

\section{Particle tracking method}
The particle tracking method, as implemented in PolyParticleTracker consists of the following steps:
\begin{enumerate}
\item Image smoothing and noise reduction,
\item Identification of particles to track and estimation of particle coordinates,
\item Subpixel refinement of particle coordinates,
\item Particle discrimination,
\item Linking of particle postions between frames.
\end{enumerate}
As with other tracking methods, it is the subpixel refinement algorithm (Step 3 above) that is the dominant factor which determines the accuracy and precision of the analysis \cite{Cheezum2001}. This is the novel part in our method, so it is described in detail below. The other steps are similar to previous methods and are described in the Appendix for completeness.

\subsection{Subpixel refinement of particle positions}
\begin{figure}
\centering
\resizebox{8cm}{!}{\includegraphics{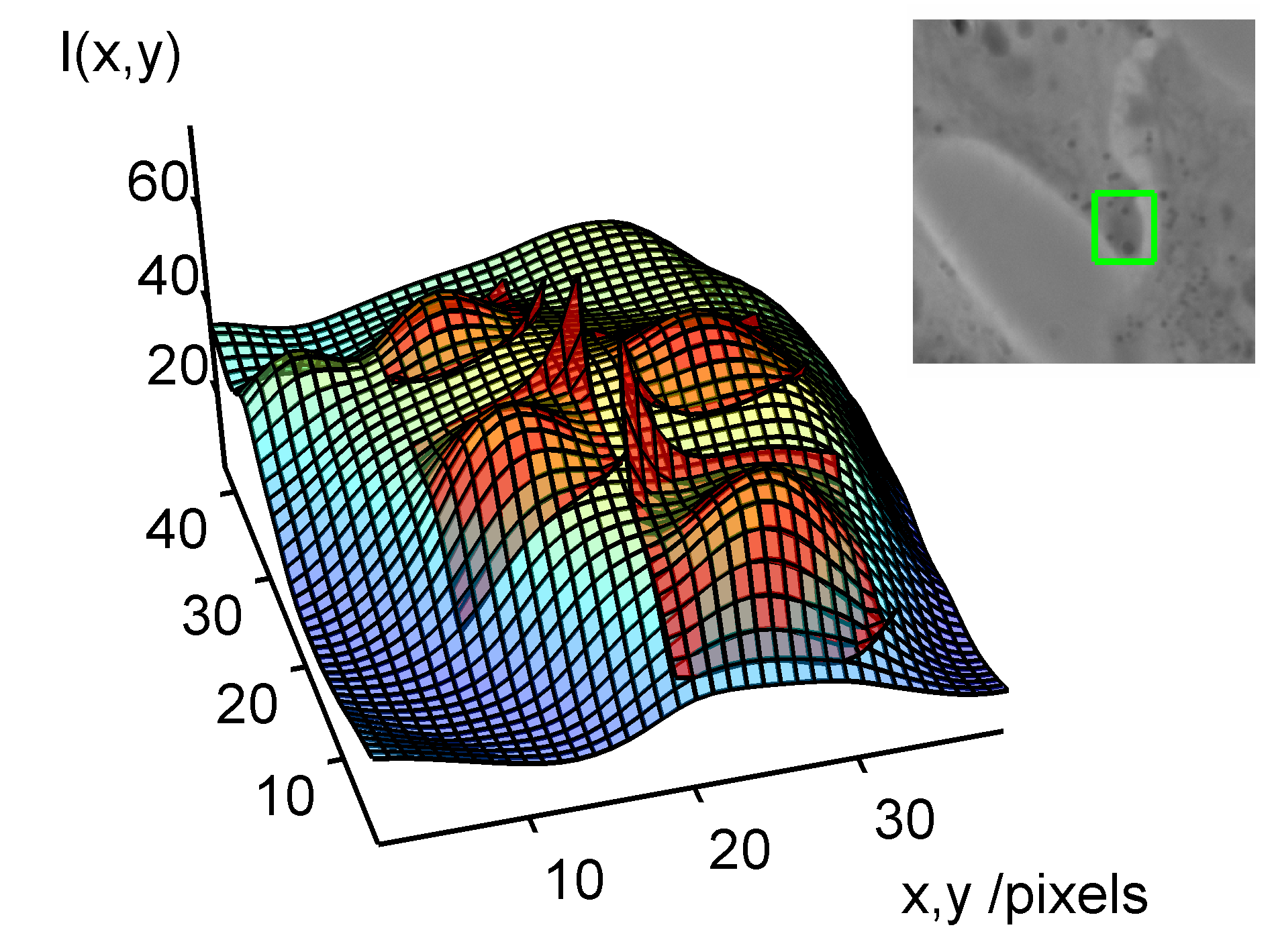}}
\caption{Example fits of particles in a cell image. The orginal image (inset) is locally fitted at each feature with a quartic surface, weighted by a Gaussian function.}
\label{fitsurf}
\end{figure}

The subpixel refinements are made as follows. For an identified particle with estimated position $(x_0,y_0)$, and estimated radius $R$, we find the closest fit (in the least-squares sense) of the intensity map of the image $I(x,y)$, with a quartic polynomial function:

\begin{equation}
I_{fit}(x,y)=\sum_{i=0,j=0}^{i+j=4} P_{ij} x^i y^j \,,
\label{ifit}
\end{equation}
where $x$ and $y$ denote pixel coordinates. The fit is given a weight at each pixel by the Gaussian function:

\begin{equation}
W(x,y)=\exp \left(-\alpha\frac{x^2+y^2}{R^2}\right) \,,
\end{equation}
where setting the decay constant $\alpha=1$ ensures that particle itself is weighted and the surroundings are excluded. The least-squares fit is sought in the domain $x_0-\beta R\le x\le x_0+\beta R$ and $y_0-\beta R\le y\le y_0+\beta R$ where setting the constant $\beta=2$ ensures that the particle is adequately encompassed within the domain. We find that the subpixel correction itself is fairly insensitive to both $\alpha$ and $\beta$. Example quartic fits of particles in a real cell image are shown in figure \ref{fitsurf}.

Having fitted the particle image, we use the coefficients of the quadratic part of $I_{fit}$ (i.e.~$P_{00}$, $P_{10}$, $P_{01}$, $P_{20}$, $P_{11}$ and $P_{02}$,) to evaluate the subpixel correction to the particle centre coordinates. The refined particle centre is defined by the extremum of the quadratic part: i.e.~we take $x_1=x_0+\Delta x$ and $y_1=y_0+\Delta y$, where

\begin{equation}
\begin{split}
\Delta x = \frac{P_{11}P_{01}-2P_{02}P_{10}}{4J} \,, \\
\Delta y = \frac{P_{11}P_{10}-2P_{20}P_{01}}{4J} \,, \\
\end{split}
\end{equation}
and $J=P_{20}P_{02}-P_{11}^2/4$. (For a reference on quadratic surfaces, see \cite{wolfram_quadric}.) However, we can make the particle centre coordinates more precise if the subpixel correction is made iteratively as follows. If either $\Delta x$ or $\Delta y$ are above a threshold fraction of a pixel $Q$, then the approximate centre is moved by $Q$, and the fitting procedure and calculation of $(\Delta x, \Delta y)$ are repeated:

\begin{equation}
\begin{split}
x_{n+1}=x_n + \left\{ \begin{array}{ll} \Delta x & \textrm{for $|\Delta x| < Q$} \\
	            \frac{\Delta x}{|\Delta x|} Q & \textrm{otherwise} \,, \end{array} \right. \,,\\
y_{n+1}=y_n + \left\{ \begin{array}{ll} \Delta y & \textrm{for $|\Delta y| < Q$} \\
	            \frac{\Delta x}{|\Delta x|} Q & \textrm{otherwise} \,, \end{array} \right. \,,
\end{split}
\end{equation}
where $\Delta x$, $\Delta y$ and the particle radius are recalculated in each iteration. Setting $Q=0.5$ gives good precision, and the cycle is ended three iterations after the conditions $\Delta x,\Delta y < Q$ are met. The cycle is also terminated if $|x_n-x_0|>\beta R$ (i.e.~the centre has wandered too far without finding the particle) or if $J<0$ (i.e.~the fit no longer resembles a particle with an extremum in intensity).

Provided that images are acquired fast enough so that particles do no move further than their own radii, from one frame to the next, then the coordinates of a particle in frame $T$ can be used as the initial estimate for its position in frame $T+1$. This procedure provides a ready method of linking a particle's trajectory (Step 5 above). The coefficients $P_{ij}$ also provide the required information to calculate the particle's radius, eccentricity, angle of rotation and skewness (see Appendix).

\section{Quantitative evaluation on simulated images}
We first reproduce the error analysis of Cheezum et al.~\cite{Cheezum2001}. Second, we introduce a new test, to simulate the tracking of a particle against a simple non-uniform background.

The analysis of Cheezum et al.~\cite{Cheezum2001} is based on simulated images of a single fluorescent spherical particle, imaged in a typical microscope with a 100x oil-immersion lens. The microscope is equipped with a CCD camera, modelled by a square array of pixels, such that the final resolution of the image is 99 nm per pixel. In the model, the particle is first represented by a circular disk of radius $R$ with uniform emission brightness on a high resolution grid of 9~nm per cell, such that 11 cells of this grid correspond to the width of a CCD pixel. The image is then broadened by the point-spread funtion:

\begin{equation}
PSF(r)=\left(\frac{2J_1(ra)}{r}\right)^2 \,,
\end{equation}
where $a=2\pi \textrm{NA} /\lambda$, NA=1.3, and $\lambda=570$~nm.  The brightness measured by each pixel of the CCD array is the average brightness of the corresponding 11x11 array of high-resolution cells. The CCD image intensity is then scaled to a peak value $S+10$ and a background value $10$, where $S$ is the signal strength and intensity is measured in number of photoelectrons. CCD images display ``shot noise'' when the photons counted by each pixel is low enough for statistical fluctuations to appear in their number. This noise follows a Poisson distribution: if the mean photoelectron count is $I$, then the standard error is $\sqrt{I}$. We used the Poisson noise function of Matlab (Natick, MA, USA) to add Poisson noise to the CCD images. The signal to noise ratio of the intensity peak is then given by $S/N = S/\sqrt{S+10}$.

Two particle radii are considered below: a ``point particle'' of 9 nm, and a larger particle of 1-$\mu$m diameter whose edge can be clearly resolved. A range of $S/N$ values are considered; at each 1000 frames are generated, in which the particle is moved by shifting the high resolution image by a whole number of cells in each of the $x$ and $y$ directions. We simulate Brownian motion by shifting the image in each direction according to a Gaussian distribution of zero mean and standard deviation 27 nm (three high-resolution cells).

The tracks generated by our method are compared with the exact track as follows. The bias and standard deviation are measured as 

\begin{equation}
\begin{split}
B = & \langle a - \hat{a} \rangle \,, \\
\sigma = & \langle (a - \langle a\rangle)^2 \rangle ^{1/2} \,,
\end{split}
\end{equation}
where the average is taken over all 1000 successive images. Here $a$ represents each of the tracked particle coordinates $x$ and $y$, and $\hat{a}$ represents the corresponding exact coordinate. 

\begin{figure}
\centering
\resizebox{8cm}{!}{\includegraphics{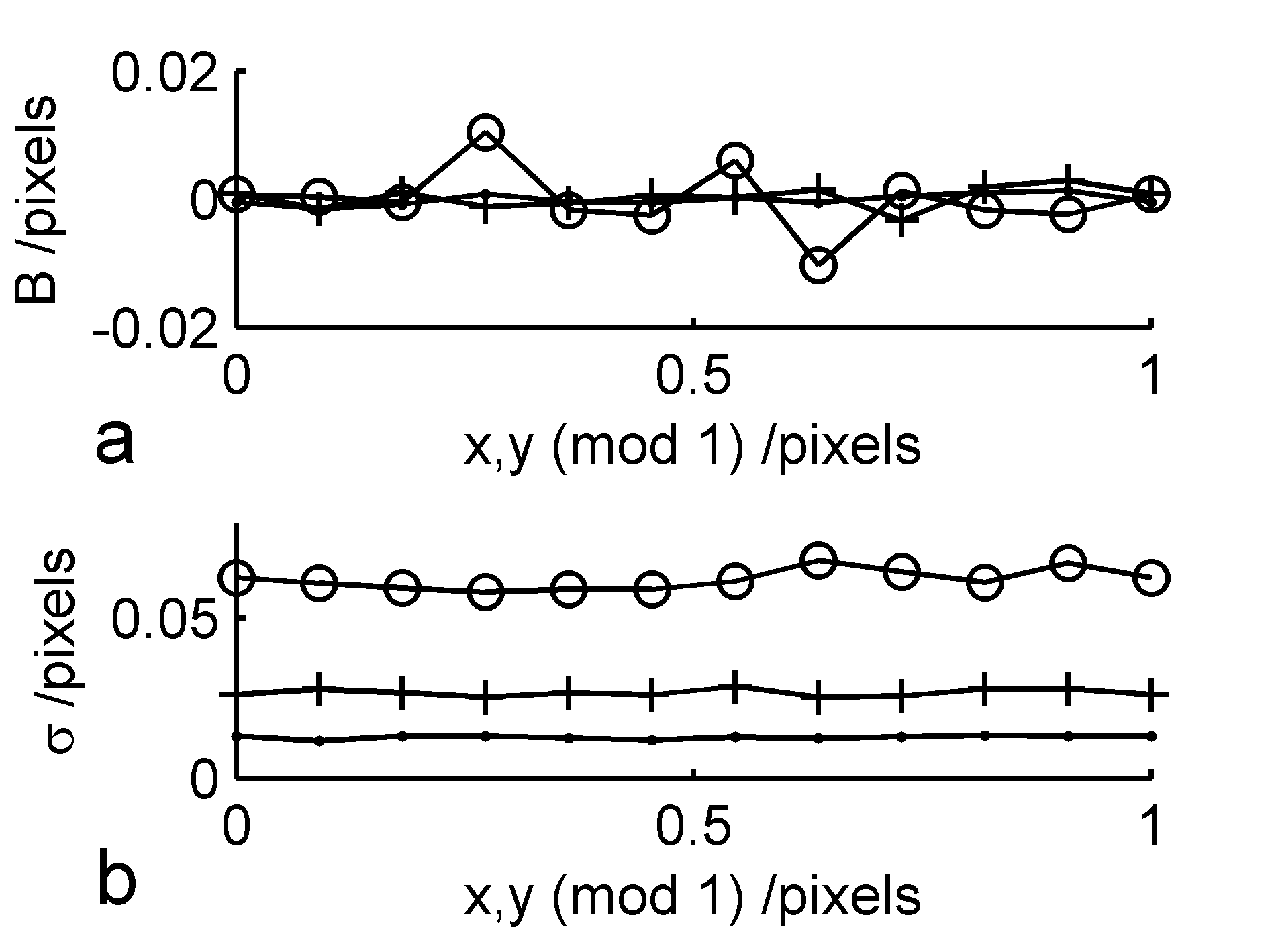}}
\caption{(a) Bias $B$ and (b) standard deviation $\sigma$ of the PFGW method measured in pixels, as a function of position in each CCD pixel, at a range of $S/N$. Both $x$ and $y$ coordinates are included simultaneously. Key: ($\circ$) $S/N=7.9$, ($+$) $S/N=16.2$, ($\cdot$) $S/N=31.3$. Both $B$ and $\sigma$ are almost uncorrelated with position, and $B$ is approximately a factor of 10 smaller than $\sigma$ at each $S/N$.}
\label{StoN1b}
\end{figure}

The results for tracking the point particle with the PFGW method are as follows: figure \ref{StoN1b} shows that both $B$ and $\sigma$ are almost uncorrelated with position relative to the pixel edges at both low and high signal strengths. Furthermore, $B$ is approximately a factor of 10 smaller than $\sigma$ at each $S/N$; thus the position of a particle relative to the CCD pixels, as determined by the method, is effectively unbiased. 

\begin{figure}
\centering
\resizebox{8cm}{!}{\includegraphics{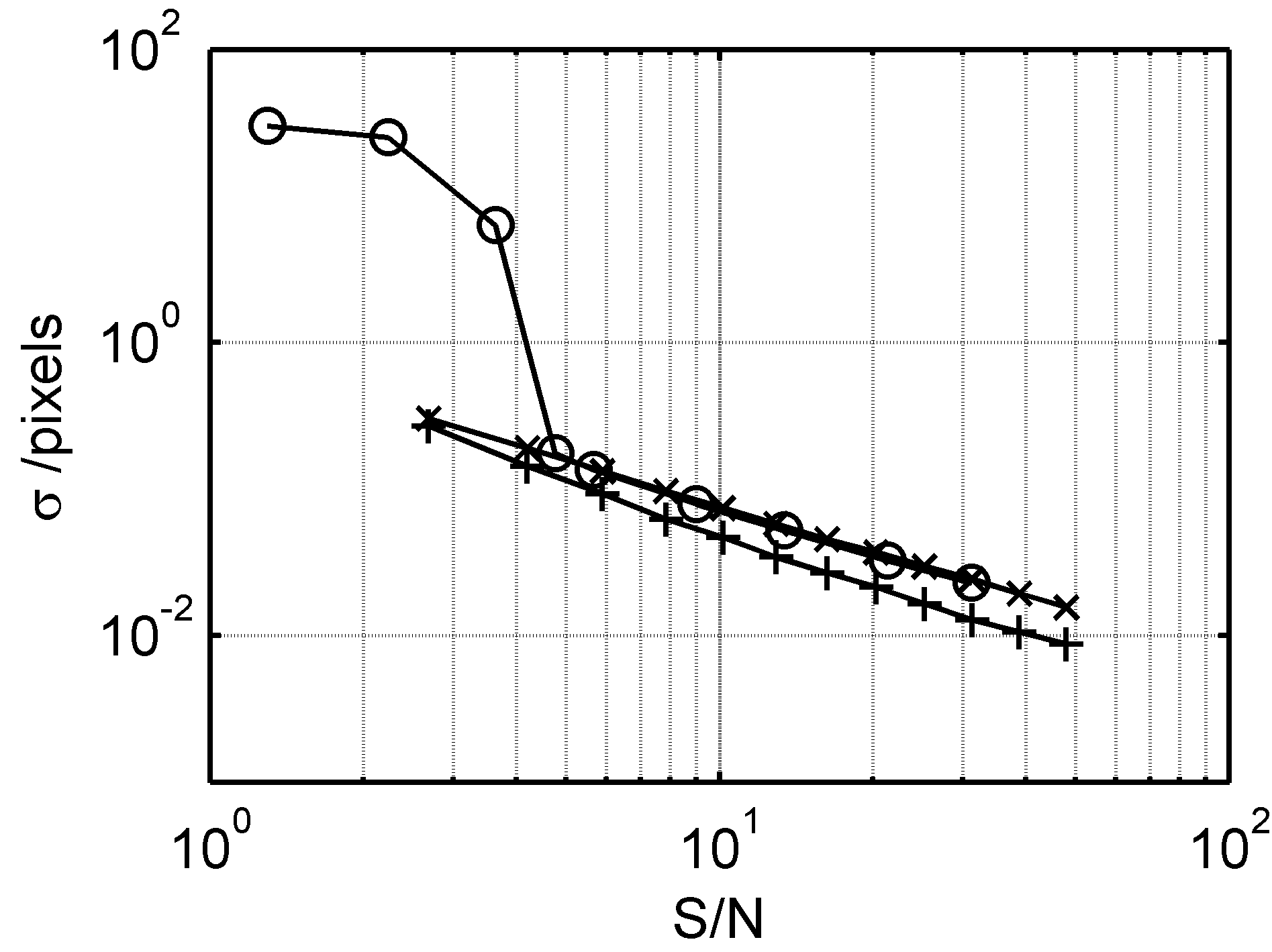}}
\caption{ Tracking precision in pixels using the PFGW method, as a function of $S/N$. The best precision is obtained for the point-particle ($+$), where $\sigma=1$~nm is reached at $S/N=41.2$. 1-$\mu$m particles are tracked with slightly higher error ($\times$). Previous results for Gaussian-fit tracking of a point ($\circ$) are plotted for comparison \cite{Cheezum2001}.}
\label{StoN2}
\end{figure}

The mean precision for tracking both the point particle and the 1-$\mu$m particle is evaluated for a range of $S/N$ and plotted in figure \ref{StoN2}. Results for Gaussian-fit tracking of a point-particle are superposed for comparison (from Cheezum et al.~\cite{Cheezum2001}). PFGW obtains better precision for the point particle than the 1-$\mu$m particle, despite the fact that more pixels are included in the fit of the latter. This may be explained by considering that the brightness profile of the 1-$\mu$m particle is almost flat near the peak---thus changes in this region make a relatively large difference to the peak position of the polynomial fit. However, it is especially encouraging that PFGW is more precise than Gaussian-fitting for the point particle, and still works well for 1-$\mu$m particles. For the former, our method reaches 1-nm precision at $S/N=41.2$.

In real images, especially images of cells in transmission light microscopy, there is usually a significant non-uniform background intensity around each particle. Even in fluorescence microscopy, there are normally more than one fluorescent particle, at various distances from each other, which can introduce new errors into their tracking.

\begin{figure}
\centering
\resizebox{8cm}{!}{\includegraphics{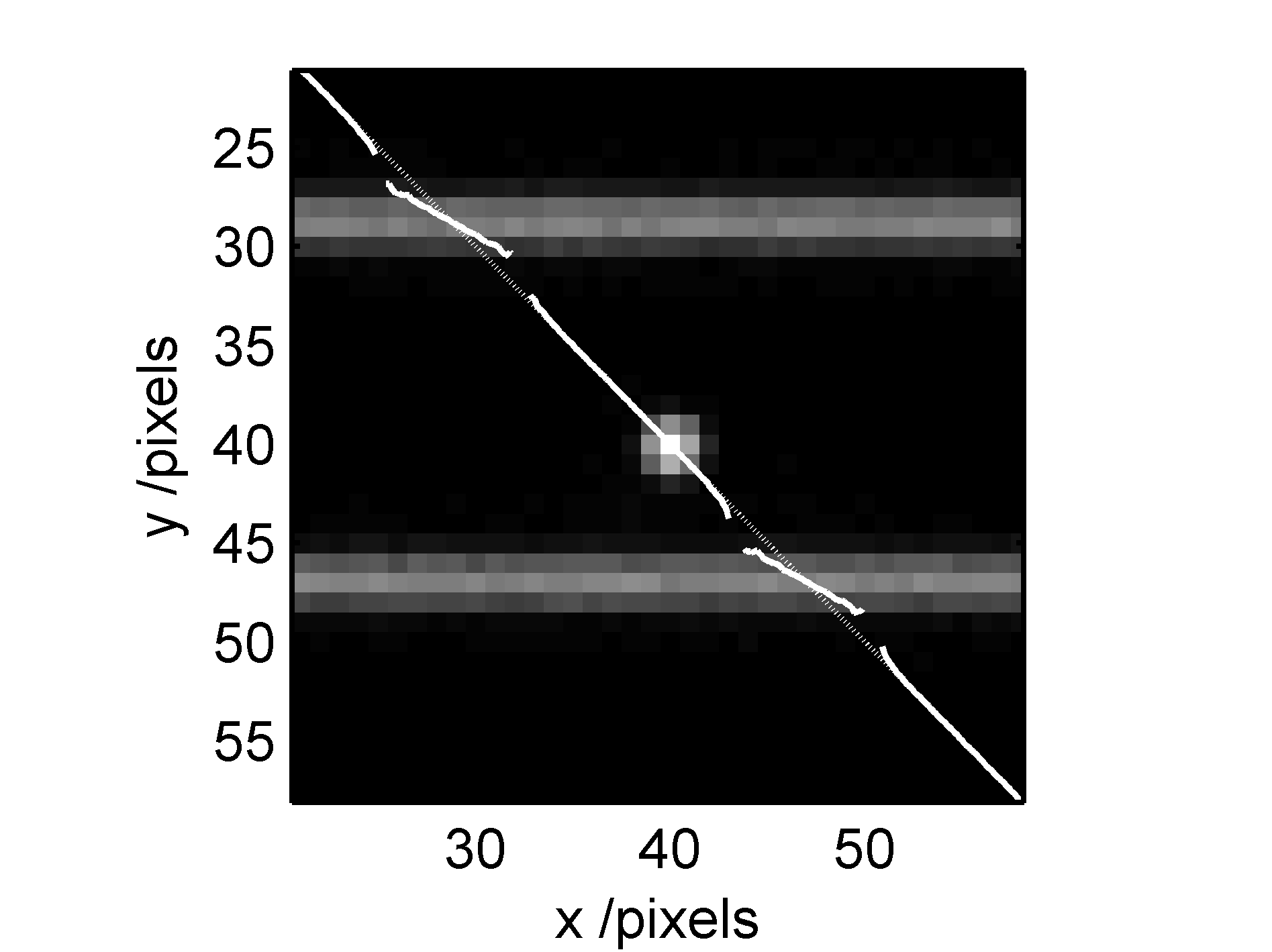}}
\caption{A simple test of particle tracking against a background: a point-particle (centre) diagonally traverses a series of thin ridges. The exact trajectory ($\cdot\cdot\cdot$) and PFGW-tracked course (---) are superposed.}
\label{ridges_image}
\end{figure}

We introduce a simple stylised background into the simulation: a series of ``ridges''---bright thin parallel lines---are placed 200 cells apart (corresponding to 18.2 CCD pixels) in a high-resolution image of the same dimensions as above. This background image is scaled such that its maximum intensity is half that of the particle image, and the two images are added. The combined image is convolved with the PSF, before producing the low-resolution CCD image and generating the noise. In 1000 successive frames, the particle is shifted diagonally across the ridges, by 27~nm in both the $x$ and $y$ directions each time. The motion was tracked using the PFGW method (figure \ref{ridges_image}). We then applied the efficient and popular method of Crocker and Grier \cite{crocker1996grier}, which is based on a centroid calculation with background subtraction, to the same images. In their method, the background is approximated by a ``boxcar average''---i.e.~a convolution of the image with a filled square of width $2w+1$ pixels. The background is then subtracted from the image, replacing with zero any intensity that falls below zero. Choosing $w$ is a compromise between two extremes: $w$ must be larger than the particle diameter, otherwise the particle itself will be subtracted, but it must not be so large that it significantly influences a particle by subtracting from it other nearby objects. Their subtraction step is particularly useful because it yields a region of zero intensity surrounding each particle, provided that the background is smoothly varying. However, the background we consider is outside this provision. 

\begin{figure}
\centering
\resizebox{8cm}{!}{\includegraphics{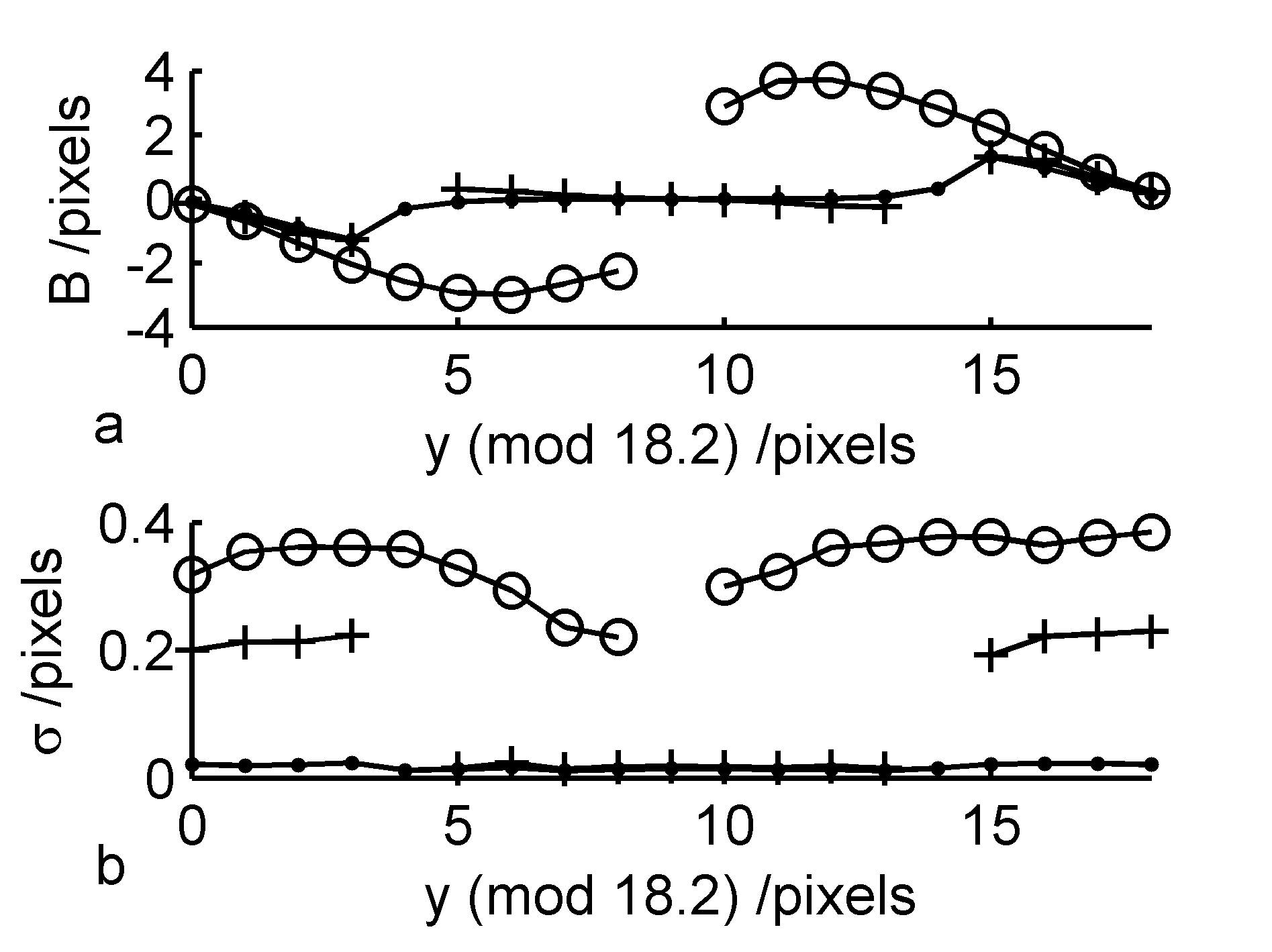}}
\caption{(a) Bias $B$ and (b) standard deviation $\sigma$ for a point-particle tracked against a background of sharp ridges, spaced periodically at 18.2 CCD pixels, where each ridge has a peak brightness of half the particle brightness, and $S/N=31.3$. Results with the PFGW method ($\cdot$) are compared with the method of Crocker and Grier \cite{crocker1996grier} where the background has been subtracted using a boxcar average of width $w=4$ ($+$) and $w=10$ ($\circ$). For the results of PFGW, $B$ is negligible when the particle and ridge images do not overlap, while $\sigma$ is unaffected by the background, even where the particle overlaps the ridge. For the Crocker-Grier method, bias and $\sigma$ are both large if the particle-ridge separation is less than $w$.}
\label{ridgebg}
\end{figure}

Figure \ref{ridgebg} shows the accuracy and precision of the point particle, as a function of its distance from one ridge to the next. The PFGW method is compared to the Crocker-Grier method, taking $w=4$ and $w=10$ in the latter. ($w=4$ just fulfils the requirement of being greater than the PSF diameter, while $w=10$ is near the maximum particle-ridge separation.) The PFGW results are biased only where the images of the particle and ridge are overlapping ($\sim$3 pixels either side of the ridge). The Crocker-Grier method results are biased within $\sim w$ pixels of the ridge, i.e.~where the boxcar function overlaps the ridge. The precision of the PFGW method is unaffected by the background (0.02 pixels throughout), but the Crocker-Grier method is much less precise where the particle is within $w$ pixels of the ridge---this is due to the noise of the background being subtracted from the particle image by the boxcar average. We note that the problems with background subtraction would be increased if the background were allowed to move, since the bias due to the background would be continually changing. This would be the case, for example, where two moving fluorescent particles approach each other.

\section{Tracking of real particles}
We demonstrate our method firstly on bright field images and secondly on images of fluorescent particles. The bright field images were taken in an Olympus IX71 inverted microscope (Olympus UK Ltd., London EC1Y 0TX, UK) with a 100x oil-immersion lens and a 1.6x magnifier, using a Photron FastCam PCI CCD camera (Photron (Europe) Ltd., Bucks SL7 1NX, UK). The fluorescent images were taken with a Leica SP2 AOBS confocal microscope (Leica Microsystems GmbH., 35578 Wetzlar, Germany). For live cell images, cells were incubated in DMEM supplemented with 10\% fetal bovine serum and 15~$\mu$m HEPES, in chambers with glass coverslip bases, incubated at 37$^\circ$C.

\begin{figure}
\centering
\subfigure[]{\resizebox{4cm}{!}{\includegraphics{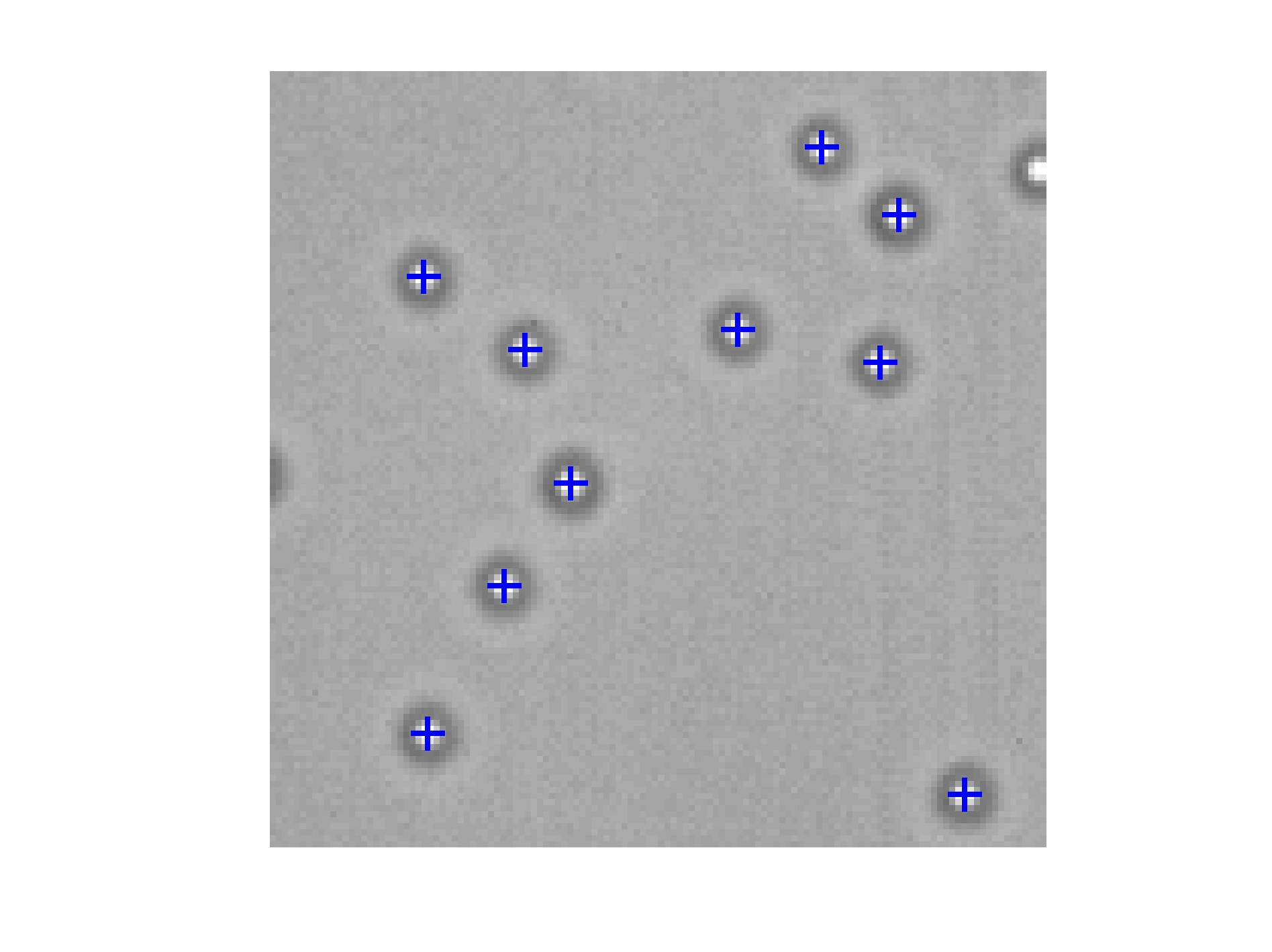}}}
\subfigure[]{\resizebox{8cm}{!}{\includegraphics{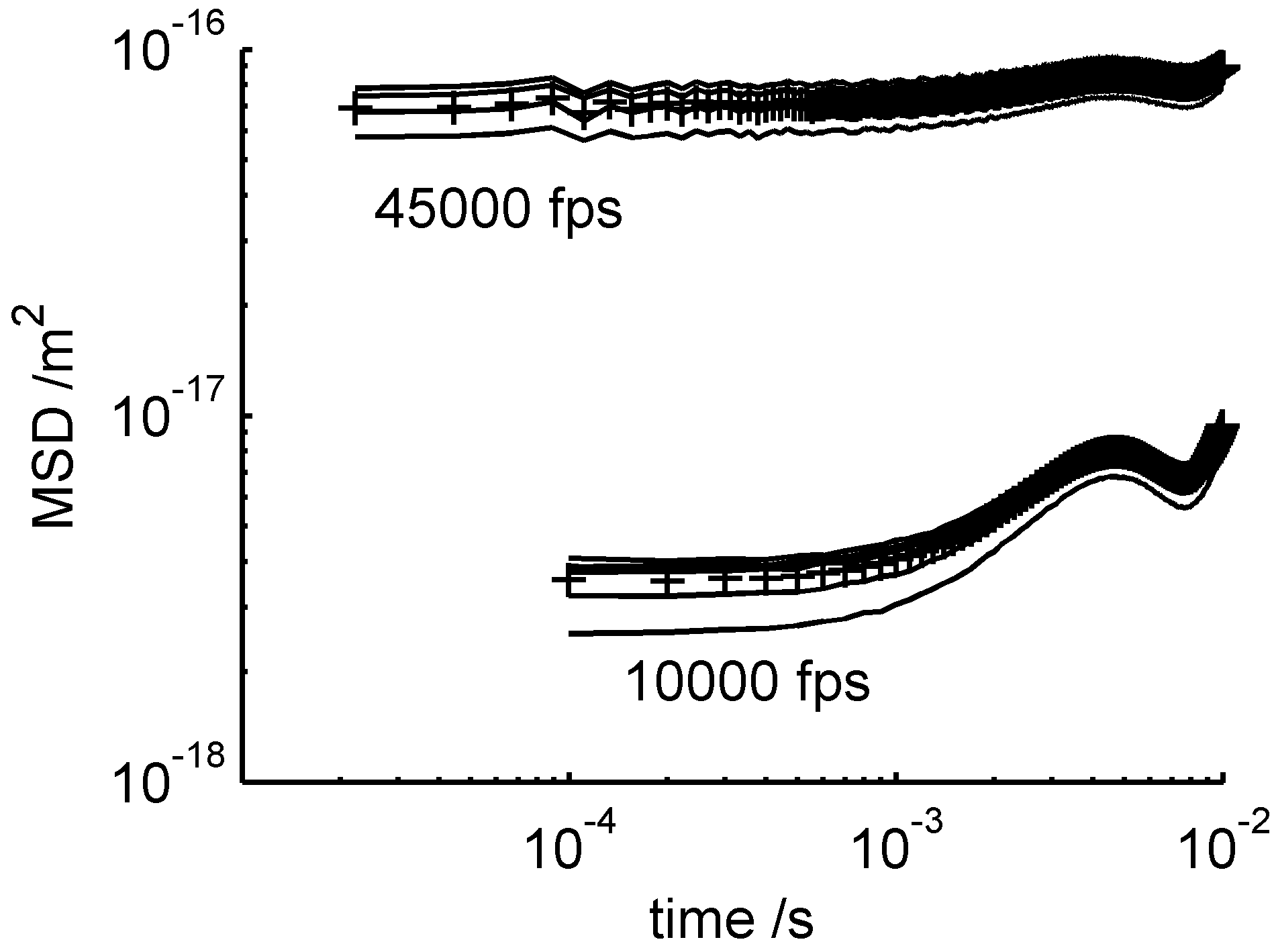}}}
\caption{(a) Sample image of immobilised 0.5-$\mu$m beads. (b) Mean square displacements at 10000 and 45000 fps. The former shows lower static error due to higher S/N.}
\label{drybeads}
\end{figure}

0.5-$\mu$m-diameter latex beads (Sigma-Aldrich, Gillingham UK, cat.~no.~L2153) were immobilised on a glass coverslip by drying, and images captured at frame rates of 10,000 and 45,000~fps. Both sets of images were captured without adjusting the illumination or focus, then particles were tracked using the PFGW method. Figure \ref{drybeads}(a) shows a sample frame, with the bead centre coordinates marked. Figure \ref{drybeads}(b) shows the mean-square displacement function (MSD), which shows a static error at short timescales of $\sqrt{\textrm{MSD}}=1.9$ and 8.3~nm for 10,000 and 45,000~fps respectively \cite{Savin2005a}. (At larger timescales a 150-Hz vibration from the building is apparent.) The reason that the static error is higher for the faster image, is that the amount of light per frame has been reduced by a factor of 4.5, and hence the signal to noise ratio has also been reduced. However, in this case, the precision is not properly described by the shot-noise model described above for two reasons: the sensitivity of the CCD pixels is far below single-photon levels, and the pixel brightness is discretized with an 8-bit colormap. For the images at 10,000 fps, we measure a mean peak pixel brightness $\langle I_{peak} \rangle = 143$ and standard deviation $(\langle (I_{peak} - \langle I_{peak} \rangle )^2 \rangle)^{1/2} = 1.15$, where the averages are taken over all time frames. These values compare with a mean brightness of the whole image $\langle I \rangle = 82.3$ and standard deviation $(\langle (I - \langle I \rangle )^2 \rangle)^{1/2} = 0.92$. Clearly the fluctuation amplitudes are approximately constant over the image, unlike shot-noise, with an amplitude of one brightness unit. Thompson et al.~\cite{Thompson2002a} have given theoretical expressions for the precision of Gaussian fitting in the two limits of shot noise and constant background noise. The latter limit (taking background noise $b=1$) predicts the present error remarkably well. By substituting our peak signal intensity of $S$ in their theory, we obtain a modified form of their equation 13~\cite{Thompson2002a}:

\begin{equation}
\sigma=\left(\frac{2 s}{\sqrt{\pi}}\right)^{1/2}\frac{b}{S} \,,
\label{thompsonmodified}
\end{equation}
where $s=1.1$ is the standard deviation of the PSF in pixels. Taking the signal to background noise ratio of the 10,000 fps data as $S/b=(143-82.3)/1$ gives $\sigma=2.0$~nm. For the 45,000 fps data, equation \ref{thompsonmodified} then predicts the standard deviation 4.5 times higher at 9.1~nm. Both values of $\sigma$ are close to the measured static errors, thus it appears that the precision is limited, in this case, by discretization of pixel brightness.

\begin{figure}
\centering
\subfigure[]{\resizebox{8cm}{!}{\includegraphics{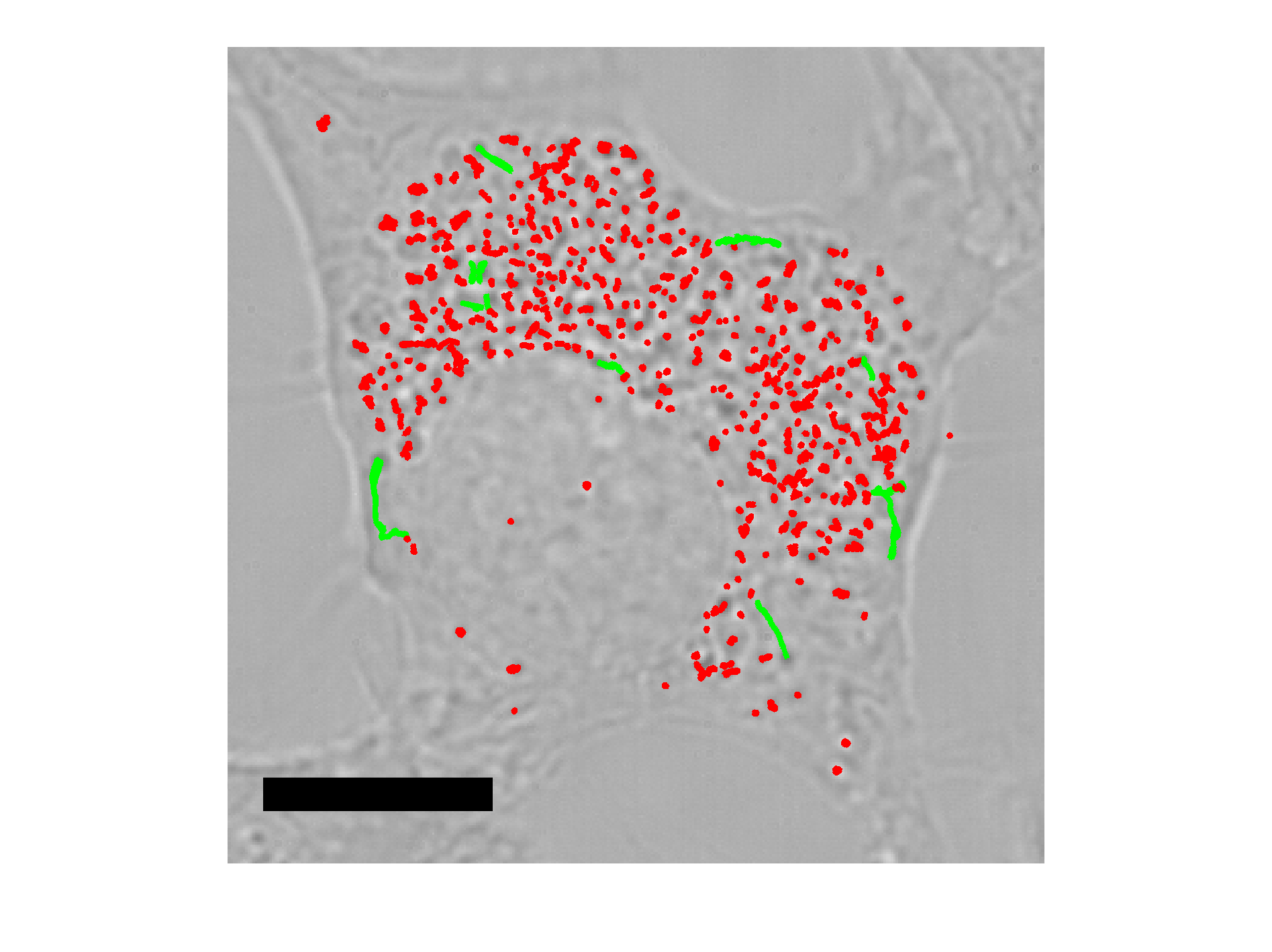}}}
\subfigure[]{\resizebox{8cm}{!}{\includegraphics{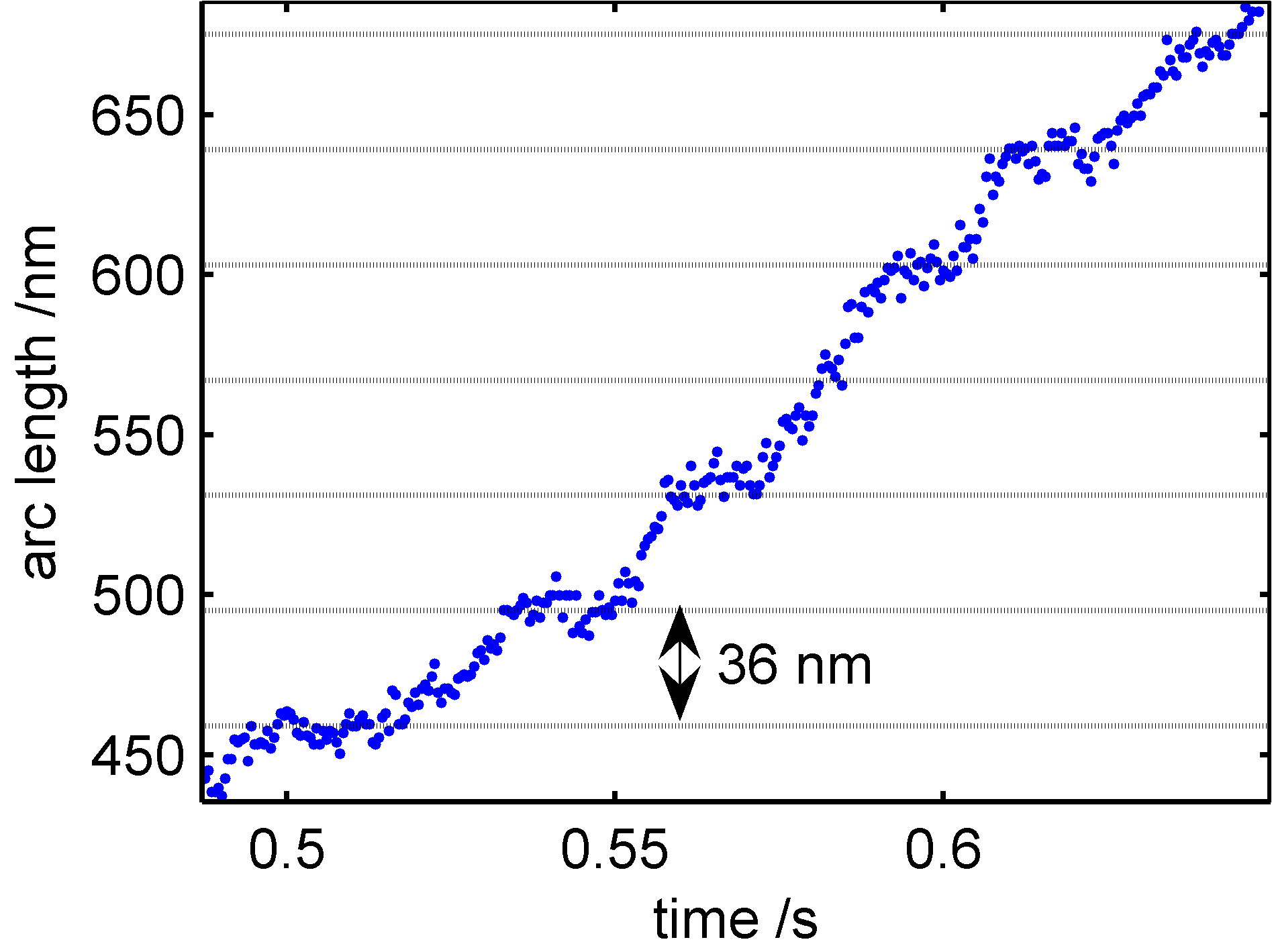}}}
\subfigure[]{\resizebox{8cm}{!}{\includegraphics{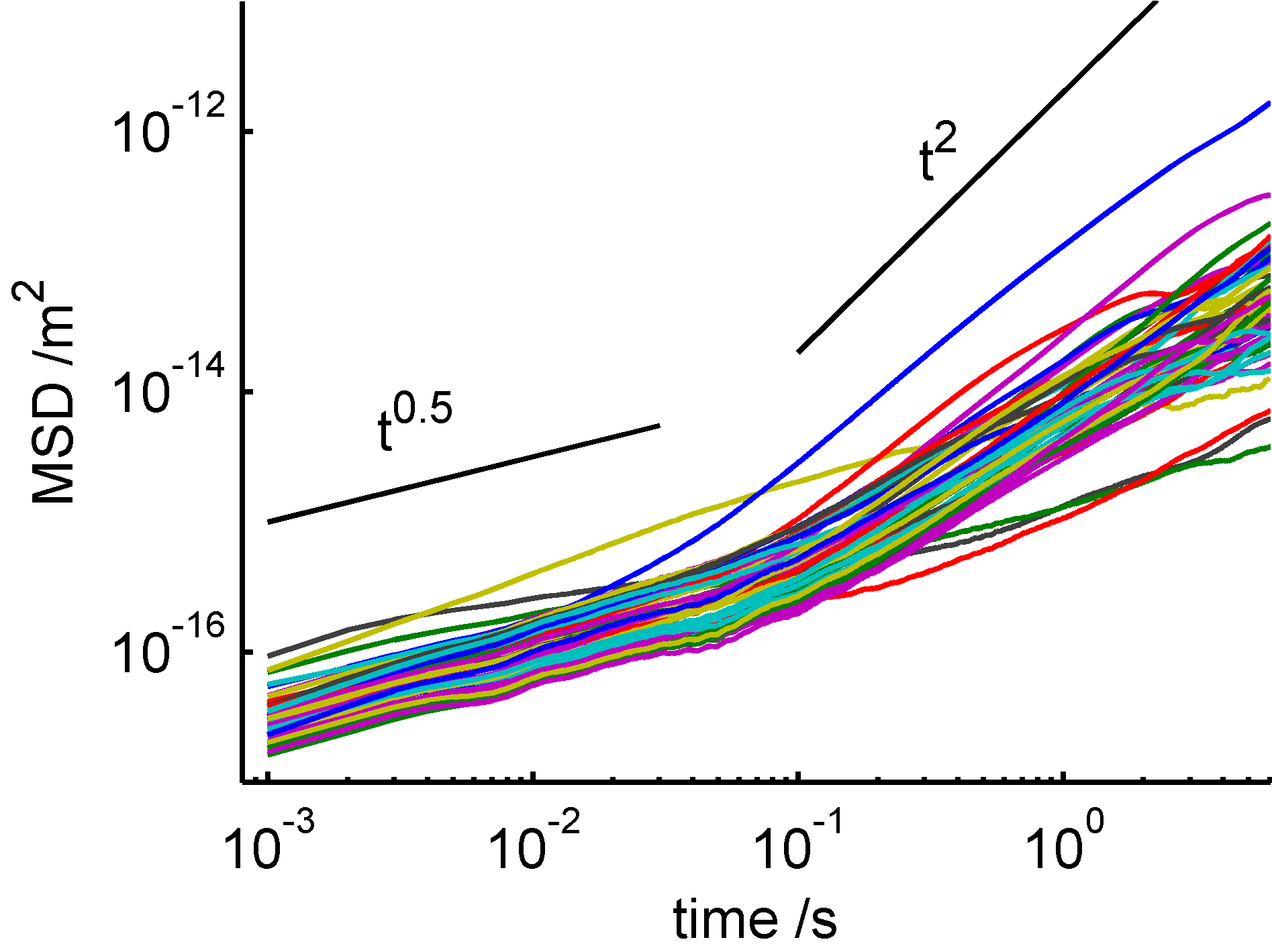}}}
\caption{(a) Tracked endogeneous particles in a cell of L929. Many particles showed active motion (highlighted green). A 10-$\mu$m scale bar is shown. (b) A run of 36-nm steps, corresponding to motion of endogeneous particles along an actin filament. (c) MSD of each particles present for more than 10,000 frames. Thermal motion (MSD$\sim t^{0.5}$) and active motion (MSD$\sim t^2$) dominate at short and long timescales respectively.}
\label{L929spread}
\end{figure}

The PFGW method can be used to track low-contrast endogeneous particles in cells. Figure \ref{L929spread}(a) shows an example of particle tracks in a well-spread L929 cell. Bright field images were captured at 1000 fps, with a final resolution of 106 nm/pixel. Some particles display only random motion, but many display active motion along tracks of various shapes (examples highlighted), some of which abruptly change direction. Although we cannot distinguish the particles or motors involved, the tracking method is of sufficient precision to see the discrete steps of many particles along the fibres of the cytoskeleton. An example run of 36-nm steps is shown in figure \ref{L929spread}(b); in this case, the particle position is plotted as an arc length along the fibre, calculated by taking the particle's nearest position along a ``smoothed'' track. The smoothed track was produced by convolving the particle's $x$ and $y$ coordinates with a Gaussian function of large width. The MSD of each particle present for more than 10,000 frames is plotted in figure \ref{L929spread}(c), and shows two different scaling regimes similar to those identified by previous authors \cite{Citters2006}. Most particles show MSD$\sim t^{1/2}$ at short times $t\lesssim 0.1$~s, indicative of tethered thermal motion on stiff filaments \cite{caspi1998}, and the exponent $\approx 2$ at long times $t\gtrsim 0.1$~s, indicative of active motion. The fact that the MSD has not reached a plateau at the fastest timescale of 1~ms shows that the measured tracks are dominated by the real motion of the particles; thus the precision is better than $\sqrt{\textrm{MSD(1 ms)}}= 4$~nm. 

\begin{figure}
\centering
\resizebox{8cm}{!}{\includegraphics{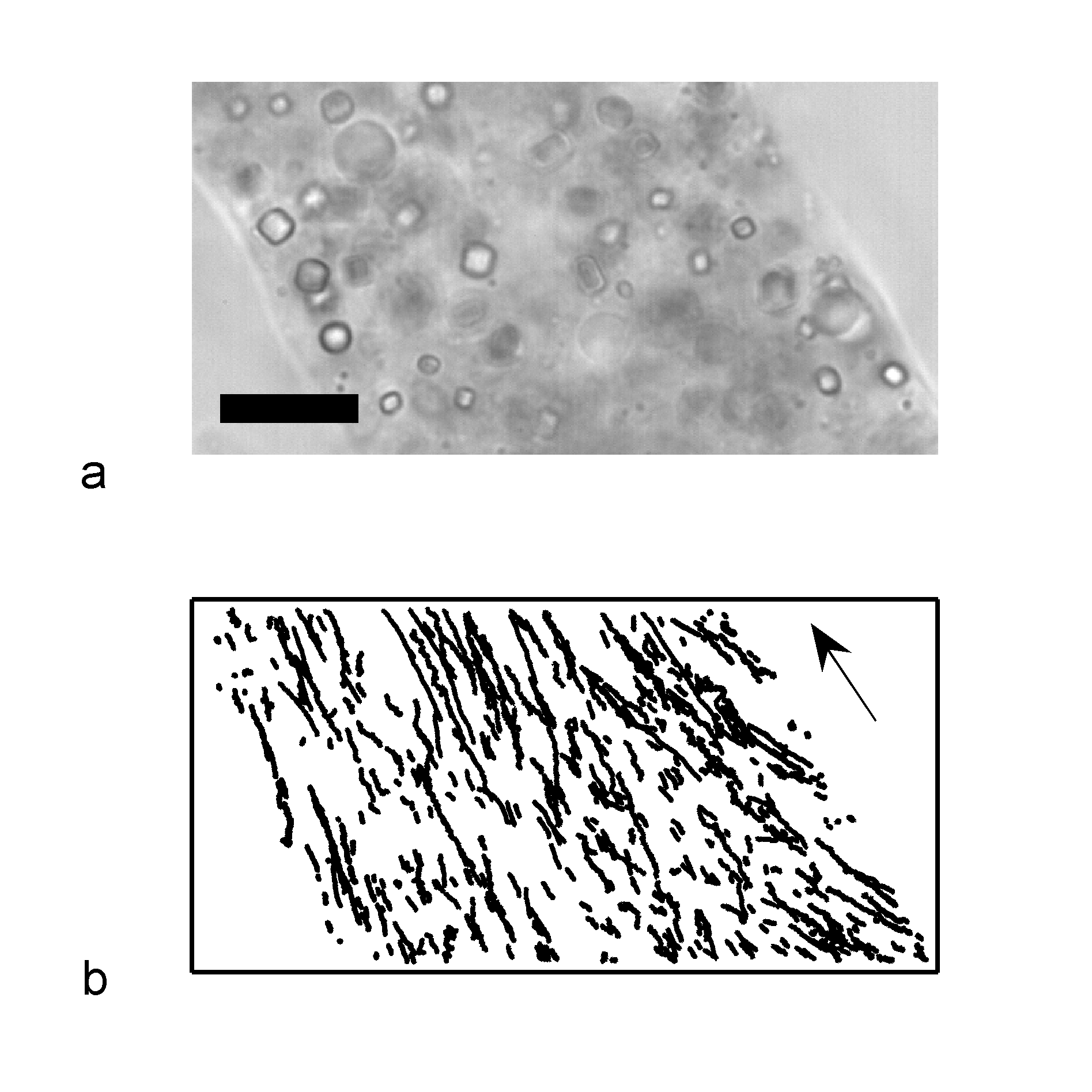}}
\caption{(a) Streaming pseudopod of \emph{A.~proteus} (10-$\mu$m scale bar shown). (b) Particles show Brownian as well hydrodynamic motion in their tracks. The direction of flow is marked with an arrow.}
\label{aprot}
\end{figure}

Another example is particle motion in streaming pseudopodia of \emph{Amoeba proteus}, as shown in figure \ref{aprot}. The presence of a large number of endogeneous particles allows its internal motion to be measured at high resolution. Our method successfully tracks particles of a variety of sizes and shapes, simultaneously.

\begin{figure}
\centering
\resizebox{8cm}{!}{\includegraphics{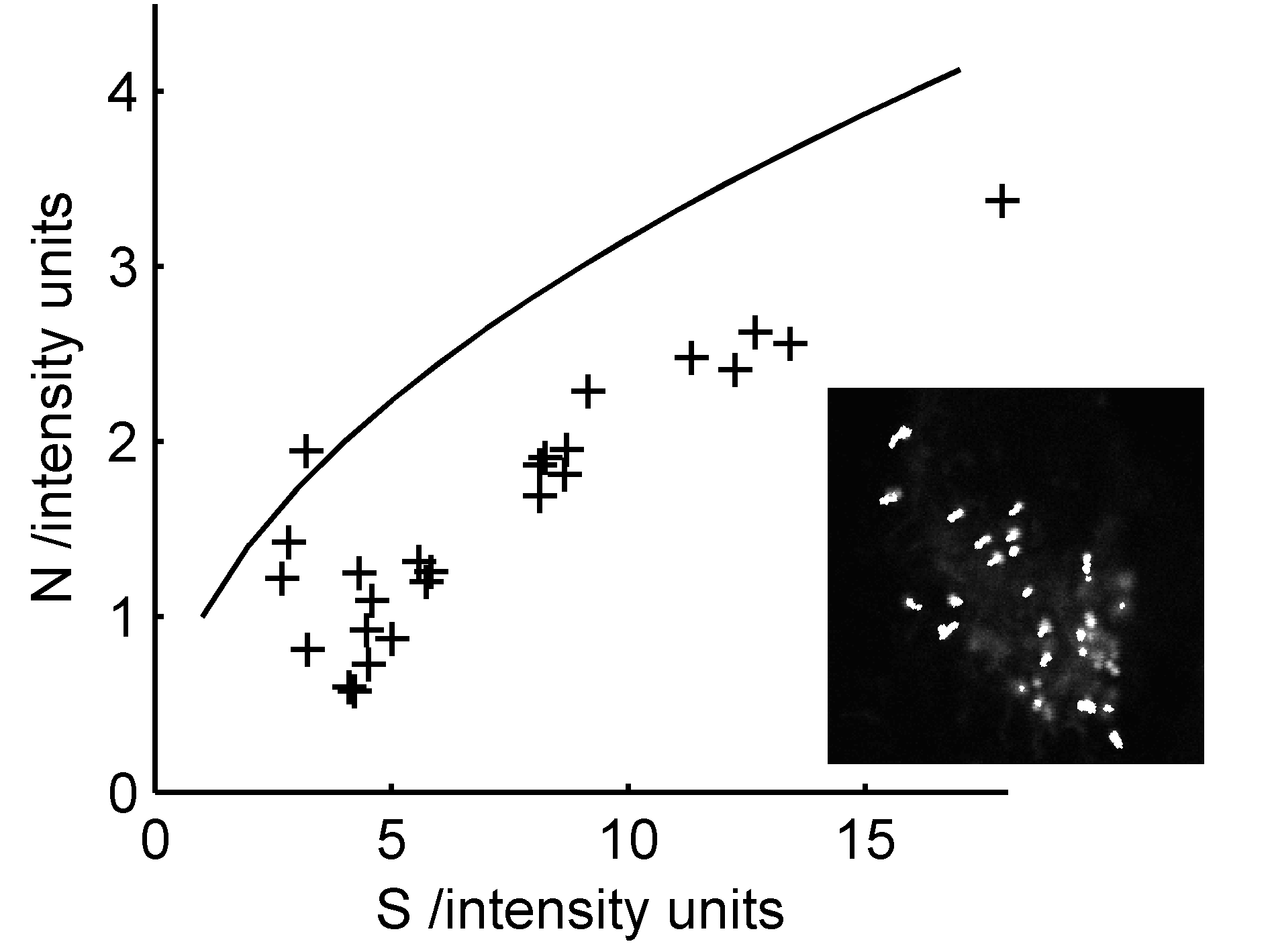}}
\caption{Tracking of fluorescent particles (inset). Peak brightness $S$ against standard deviation of fluctuations $N$ ($+$), shows reasonable correspondence with the shot-noise model, displayed here by the line $N=\sqrt{S}$ (---). The PFGW model was successful at tracking particles of $S/N \gtrsim 3$.}
\label{fluorescence}
\end{figure}

Finally we used fluorescence microscopy to track motion in HeLa cells of the same latex beads mentioned above, which fluoresced in green light. The beads were endocytosed by the cells: figure \ref{fluorescence} (inset) shows a portion of a cell with bead tracks superposed. The brightness fluctuations at each particle peak were measured and show a reasonable correspondence to the shot noise model in figure \ref{fluorescence}; each peak brightness $S$ is plotted against standard deviation of fluctuations $N$. The PFGW method successfully tracked particles with $S/N \gtrsim 3$.

We are currently applying the PFGW method to the microrheology and particle imaging velocimetry of cortical oscillations and amoeboid migration (article in preparation).

\section{Conclusion}
Our particle tracking method, based on polynomial fitting of feature points with a Gaussian weighting function, is distinguished from previous methods of particle tracking by its inherant ability to track particles against a complex background, while maintaining a similar precision to Gaussian fitting at a range of signal to noise ratios. It is ideal for particle tracking in situations where a complicated background or neighbouring particles produce errors in other methods, e.g.:~intracellular microrheology and tracking in dense colloidal situations.

\subsection*{Instructions for obtaining PolyParticleTracker}
The package PolyParticleTracker can be downloaded from our website: \\ http://personalpages.manchester.ac.uk/staff/salman.rogers/polyparticletracker/. \\
The package includes a graphical user interface, and requires Matlab version 7.0 or higher.

\subsection*{Acknowledgements}
This project was funded by the UK EPSRC under grant no.~EP/E013988/1. We gratefully acknowledge Peter March for assistance with the confocal microscopy experiments, and Marcus Jahnel and Aris Papagiannopoulos for many fruitful discussions. L929 cells were a kind gift of Ewa Paluch.

\appendix
\section{Details of PolyParticleTracker}

\subsection{Image smoothing and noise reduction}
Following Crocker and Grier \cite{crocker1996grier}, we smooth each image to correct for discretization noise, by convolving the image with a Gaussian function:

\begin{equation}
I(x,y)=\sum_{i=-w}^w\sum_{j=-w}^w I_{raw}(x+i,y+j) \exp\left(-\frac{i^2+j^2}{4\lambda^2}\right)\,,
\end{equation}
where the correlation length $\lambda$ is set to 1, and setting $w=3\lambda$ sufficiently approximates the unbounded convolution. Note the background is not subtracted as in previous particle tracking methods.

\subsection{Identification of particles to track and estimation of particle coordinates}
Particles are identified by the presence of either a local maximum or minimum in $I(x,y)$. For each of these extrema $(x_0,y_0)$, the adjacent points of inflexion are found along the orthogonal lines defined by $(x,y_0)$ and $(x_0,y)$. The mean of the distances to the four points of inflexion is used as the initial radius estimate $R_0$. $x_0,y_0,R_0$ are then input into the subpixel correction routine, described above.

\subsection{Particle discrimination}
For each particle, the intensity map $I(x,y)$ and the parameters of the polynomial fit $P_{ij}$ allow calculation of the particle's eccentricity, rotation, radius, average brightness and skewness, as follows.

Eccentricity $e$ and rotation $\theta$, i.e.~the angle between the major axis of the particle's brightness peak and the $x$ axis, are both obtained by considering the quadratic terms which dominate the vicinity of the peak. These terms define a surface whose horizontal section is an ellipse: i.e.~$\sum_{i=0,j=0}^{i+j=2} P_{ij} x^i y^j = 0$. Hence \cite{wolfram_ellipse}

\begin{equation}
\begin{split}
\theta = & \frac{1}{2}\cot^{-1}\frac{P_{20}-P_{02}}{P_{11}}\,, \\
e = & \sqrt{ 1 + \frac{\sqrt{(P_{20}-P_{20})^2 +P_{11}^2} + (P_{20}+P_{20})}
					{\sqrt{(P_{20}-P_{20})^2 +P_{11}^2} - (P_{20}+P_{20})} } \,.
\end{split}
\end{equation}

\begin{figure}
\centering
\resizebox{8cm}{!}{\includegraphics{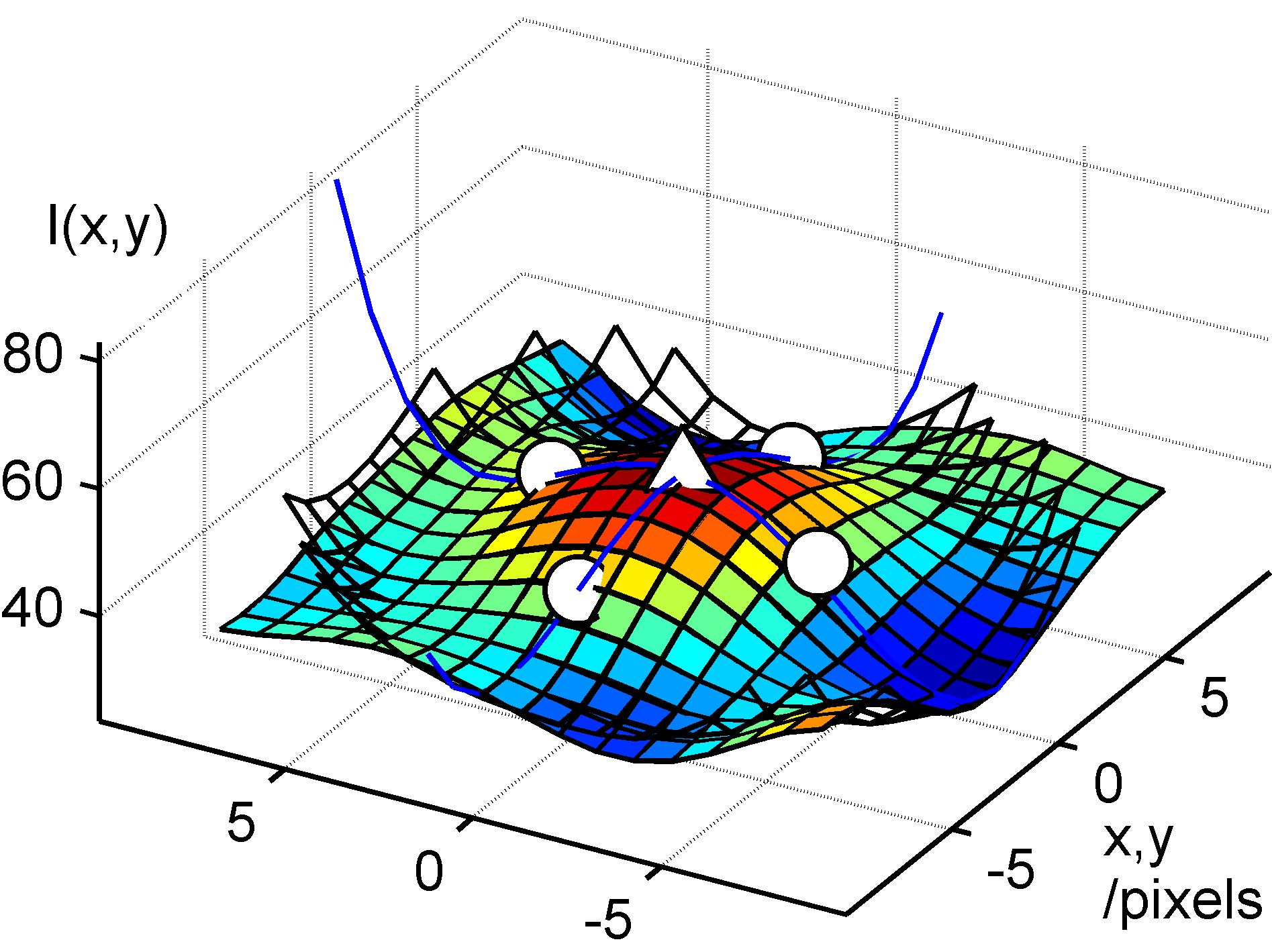}}
\caption{Example estimation of particle radius. $I(x,y)$ is represented by the solid surface, while $I_{fit}(x,y)$ is the transparent mesh. The peak of $I_{fit}(x,y)$ is marked by a triangle and its four points of inflexion in the directions $\theta,\theta+\pi/2,\theta+\pi,\theta+3\pi/2$ are marked by circles. The estimated radius is the mean distance from the peak to the points of inflexion.}
\label{radiuscalc}
\end{figure}

Calculation of the particle's radius $R$ is more complicated, as the true shape of the particle is not known, and the background intensity is non-negligible. However, a good estimate of $R$ is given by the geometric mean of the distances from the peak of $I_{fit}(x,y)$ to the four points of inflexion in the directions of the particle's major and minor axes, i.e.~the directions $\theta,\theta+\pi/2,\theta+\pi,\theta+3\pi/2$. The radius calculation is illustrated in figure \ref{radiuscalc}. In terms of the polynomial coefficients:

\begin{equation}
R=\left(\frac{A B}{36 C D}\right)^{1/4} \,,
\end{equation}
where
\begin{equation}
\begin{split}
A = & P_{20}\cos^2\theta - P_{11}\cos\theta\sin\theta + P_{02}\sin^2\theta \,, \\
B = & P_{20}\sin^2\theta - P_{11}\cos\theta\sin\theta + P_{02}\cos^2\theta \,, \\
C = & P_{40}\cos^4\theta - P_{31}\cos^3\theta\sin\theta + P_{22}\cos^2\theta\sin^2\theta - P_{13}\cos\theta\sin^3\theta + P_{04}\sin^4\theta \,, \\
D = & P_{40}\sin^4\theta + P_{31}\sin^3\theta\cos\theta + P_{22}\sin^2\theta\cos^2\theta + P_{31}\sin\theta\cos^3\theta + P_{04}\cos^4\theta \,. \\
\end{split}
\end{equation}

The cubic terms in $I_{fit}(x,y)$ determine how skewed or lopsided the particle peak is. We can define an approximate \emph{skewness} of a particle by analogy to the skewness of a statistical distribution, as the ratio of the cubic terms to the quadric terms at $x,y\approx R$:

\begin{equation}
\textrm{Skewness} = \frac{ |P_{30}| + |P_{21}| + |P_{12}| + |P_{03}| }{ P_{20}P_{02}-P_{11}^2/4 } R \,.
\end{equation}

The eccentricity, brightness, radius and skewness can all be used to discriminate whether a particle is acceptable for tracking or not. 

\subsection{Linking of particle postions between frames}
As mentioned above, the coordinates of a particle in frame $T$ are used as the initial estimate for its position in frame $T+1$, providing a reliable method of linking a particle's trajectory, provided the particle does not move further than its own radius, from $T$ to $T+1$. If, however, the particle has moved further, PolyParticleTracker will look for a nearby extremum in $I(x,y)$, which is nearer than the next-nearest identified particle and has a similar brightness to the original particle in frame $T$. PolyParticleTracker can also be set to search for new particles every $N$ frames.

\bibliographystyle{unsrt}
\bibliography{ssrbib-manchester}

\begin{thebibliography}{10}

\bibitem{Tseng2002}
Y.~Tseng, T.~P. Kole, and D.~Wirtz.
\newblock Micromechanical mapping of live cells by multiple-particle-tracking
  microrheology.
\newblock {\em Biophys J}, 83(6):3162--3176, Dec 2002.

\bibitem{Weihs2006}
D.~Weihs, T.~G. Mason, and M.~A. Teitell.
\newblock Bio-microrheology: a frontier in microrheology.
\newblock {\em Biophys J}, 91(11):4296--4305, Dec 2006.

\bibitem{waigh2006}
T.~A. Waigh.
\newblock Microrheology of complex fluids.
\newblock {\em Reports on Progress in Physics}, 68(3):685--742, 2005.

\bibitem{Bausch1998}
A.~R. Bausch, F.~Ziemann, A.~A. Boulbitch, K.~Jacobson, and E.~Sackmann.
\newblock Local measurements of viscoelastic parameters of adherent cell
  surfaces by magnetic bead microrheometry.
\newblock {\em Biophys J}, 75(4):2038--2049, Oct 1998.

\bibitem{Ashkin1997}
A.~Ashkin.
\newblock Optical trapping and manipulation of neutral particles using lasers.
\newblock {\em Proc Natl Acad Sci U S A}, 94(10):4853--4860, May 1997.

\bibitem{Courty2006a}
S.~Courty, C.~Luccardini, Y.~Bellaiche, G.~Cappello, and M.~Dahan.
\newblock Tracking individual kinesin motors in living cells using single
  quantum-dot imaging.
\newblock {\em Nano Lett}, 6(7):1491--1495, Jul 2006.

\bibitem{Watanabe2007a}
T.~M. Watanabe and H.~Higuchi.
\newblock Stepwise movements in vesicle transport of her2 by motor proteins in
  living cells.
\newblock {\em Biophys J}, 92(11):4109--4120, Jun 2007.

\bibitem{adrian2005}
R.J. Adrian.
\newblock Twenty years of particle image velocimetry.
\newblock {\em Experiments in Fluids}, 39:157--483, 2005.

\bibitem{grant1997}
I.~Grant.
\newblock Particle image velocimetry: a review.
\newblock {\em Proceedings of the Institution of Mechanical Engineers, Part C:
  Journal of Mechanical Engineering Science}, 211:55--76, 1997.

\bibitem{Miura2005}
K.~Miura.
\newblock Tracking movement in cell biology.
\newblock {\em Adv Biochem Eng Biotechnol}, 95:267--295, 2005.

\bibitem{Cheezum2001}
M.~K. Cheezum, W.~F. Walker, and W.~H. Guilford.
\newblock Quantitative comparison of algorithms for tracking single fluorescent
  particles.
\newblock {\em Biophys J}, 81(4):2378--2388, Oct 2001.

\bibitem{crocker1996grier}
J.~C. Crocker and D.~G. Grier.
\newblock Methods of digital video microscopy for colloidal studies.
\newblock {\em Journal of Colloid and Interface Science}, 179(1):298--310,
  1996.

\bibitem{Sbalzarini2005}
I.~F. Sbalzarini and P.~Koumoutsakos.
\newblock Feature point tracking and trajectory analysis for video imaging in
  cell biology.
\newblock {\em J Struct Biol}, 151(2):182--195, Aug 2005.

\bibitem{Carter2005}
B.~C. Carter, G.~T. Shubeita, and S.~P. Gross.
\newblock Tracking single particles: a user-friendly quantitative evaluation.
\newblock {\em Phys Biol}, 2(1):60--72, Mar 2005.

\bibitem{wolfram_quadric}
http://mathworld.wolfram.com/quadraticsurface.html.

\bibitem{Savin2005a}
T.~Savin and P.~S. Doyle.
\newblock Static and dynamic errors in particle tracking microrheology.
\newblock {\em Biophys J}, 88(1):623--638, Jan 2005.

\bibitem{Thompson2002a}
R.~E. Thompson, .D~R. Larson, and W.~W. Webb.
\newblock Precise nanometer localization analysis for individual fluorescent
  probes.
\newblock {\em Biophys J}, 82(5):2775--2783, May 2002.

\bibitem{Citters2006}
K.~M.~Van Citters, B.~D. Hoffman, G.~Massiera, and J.~C. Crocker.
\newblock The role of f-actin and myosin in epithelial cell rheology.
\newblock {\em Biophys J}, 91(10):3946--3956, Nov 2006.

\bibitem{caspi1998}
A.~Caspi, M.~Elbaum, R.~Granek, A.~Lachish, and P.~Zbaida.
\newblock Semi-flexible polymer network: a view from inside.
\newblock {\em Physics Review Letters}, 80:1106--1109, 1998.

\bibitem{wolfram_ellipse}
http://mathworld.wolfram.com/ellipse.html.

\end{thebibliography}

\end{document}